\documentclass[a4paper]{article}

\usepackage{amsfonts, amssymb}

\newtheorem{theo}{Theorem}

\newtheorem{coro}{Corollary}
\newtheorem{prop}{Proposition}
\newtheorem{lemm}{Lemma}

\newcommand{\pr}{\indent{\em Proof: \ }}
\newcommand{\qed}{\hspace*{5 mm}$\square$\bigskip}
\newenvironment{proof}{\noindent {\pr}\ }{\qed}

\newcommand{\Z}{{\mathbb{Z}}}
\newcommand{\C}{{\cal C}}
\newcommand{\zero}{{\mathbf{0}}}
\newcommand{\n}{{N(u_0)}}
\newcommand{\vx}{{\{v_1,v_2,v_3,v_4,v_5,v_6\}}}
\newcommand{\trip}{{\{v_1,v_2,v_3\}}}

\title{On the minimum distance graph of an extended Preparata code }

\author{C. Fern{\'a}ndez-C{\'o}rdoba \thanks{C. Fern{\'a}ndez-C{\'o}rdoba is at Department of Information
and Communications Engineering, Universitat Aut\`{o}noma de Barcelona, 08193-Bellaterra, Spain. Author wishes to
acknowledge the joint sponsorship of the Fulbright Program in Spain and the Ministry of Science and Innovation
during her research stay at Auburn University.}  \and K. T. Phelps \thanks{K. T. Phelps is at
              Mathematics and Statistics Department, Auburn University, Auburn 36849 AL, USA.}}
\date{}

\begin{document}

\maketitle

\begin{abstract}
The minimum distance graph of an extended Preparata code $P(m)$ has vertices
 corresponding to codewords and edges corresponding to pairs of codewords that are distance 6 apart.
 The clique structure of this graph  is investigated and  it is established that the minimum
 distance graphs of two extended Preparata
codes are isomorphic if and only if the codes  are equivalent.
\end{abstract}

%%%%%%%%%%%%%%%%%%%%%%%%%%%%
\section{Introduction}
%%%%%%%%%%%%%%%%%%%%%%%%%%%%

Let $\Z_2^n$ be the $n$-dimensional binary vector space. The (Hamming) distance between two vectors
$x,y\in\Z_2^n$ is the number of coordinates in which they differ and it is denoted by $d(x,y)$. The weight of a
vector $x\in\Z_2^n$ is the number of its non-zero entries. Let $\C\subseteq \Z_2^n$ be a (binary) code of length
$n$ and $c\in \C$ a codeword. The minimum distance of $\C$ is the minimum distance of its codewords. The
support of $c$, denoted as $supp(c)$, is the set of non-zero coordinates of $c$. Two codes $\C_1$ and $\C_2$ are
isometric if there exists a one-to-one map $I:\C_1\rightarrow \C_2$, such that $d(x,y)=d(I(x),I(y))$, for all
$x,y\in\C_1$. Two codes are called equivalent if one can be obtained from the other by translation
and permutation of coordinates.

A $1$-perfect code $\C$ of length $n$ is a code such that for any vector $v\in\Z_2^n$ there is a unique codeword
$c\in \C$ at a distance of at most $1$ of $v$. An extended $1$-perfect code is obtained from a $1$-perfect code by
adding a parity check coordinate. The minimum distance of a $1$-perfect code is $3$ whereas the minimum distance
of an extended $1$-perfect code is $4$.

Let $\C$ be a code of length $n$ and minimum distance $d$. The
minimum distance graph $DG$ of $\C$ is the graph whose vertices are
the codewords of $\C$, where two vertices are adjacent if and only
if the corresponding codewords are at distance $d$ apart. We will
write the vertices of $DG$ as the support set of the corresponding
codewords. Conversely, for every vertex
$v=\{v_1,\dots,v_s\}\subseteq\{1,\dots,n\}$, we will denote the
corresponding codeword as $c(\{v_1,\dots,v_s\})$. A clique in a
graph $G$ is a set of vertices of $G$ such that every pair of
these vertices are adjacent in $G$. For $t\subseteq\{1,\dots,n\}$, we denote $C(t)$ the clique
in $DG$ such that $t=\bigcap v$, for all $v$ in the clique.

A $t-(n,k,\lambda)$ design is a set of  $n$ points $V$ and a collection, $B$, of $k$-tuples called blocks, such
that any $t$-tuple of elements of $V$ is in exactly $\lambda$ blocks. A $2-(n,3,1)$ design is a called a Steiner
triple system and a $3-(n,4,1)$ is called a Steiner quadruple system. If $V$ is the set $\{1,\dots,n\}$, then we
denote a Steiner triple system as $STS(n)$ and a Steiner quadruple system as $SQS(n)$. In a $SQS(n)$, any triple is included in exactly one block
of $B$. Given a $SQS(n)$ and a triple $\{i,j,k\}$, we define $X(\{i,j,k\})$ as the element in $\{1,\dots,n\}$
such that $\{i,j,k,X(\{i,j,k\})\} \in B$. If we consider a code
$\C$, then the support of codewords of the same weight may have a design structure. In \cite{AM} we can find
sufficient conditions for the supports of the codewords of the same weight to form a $t$-design. In particular,
for any $1$-perfect code of length $2^m-1$, the codewords of weight $3$ form a $STS(2^m-1)$ and the codewords of weight
$4$ of any extended $1$-perfect code of length $2^m$ form a $SQS(2^m)$.

The Preparata code is a nonlinear distance invariant code of
length $2^m-1$, $m$ even, $m\geq 4$, and minimum distance $5$
\cite{MacW}. We denote as $P(m)$, the extended Preparata code of length $n=2^m$, obtained from the Preparata code by adding a parity check coordinate.
Note that $P(m)$ has minimum distance $6$. In \cite{ZZS} it was
shown that any extended Preparata code is a subcode of an extended
$1$-perfect code of length $n$. We denote $\C_{P(m)}$ the extended
perfect code containing the extended Preparata code $P(m)$.

\begin{lemm}\label{lemma:blockInPrep}
Let $P(m)$ be an extended Preparata code and $\C_{P(m)}$ the extended $1$-perfect code
containing it. Then, if $c$ is a codeword of $\C_{P(m)}$ of weight
$4$, there is no codeword $c'$ in $P(m)$ of weight $6$ such that
$supp(c)\subset supp(c')$.
\end{lemm}

\begin{proof}
Otherwise, $c,c'\in \C_{P(m)}$ and $d(c,c')=2$.
\end{proof}

\begin{theo}[\cite{ZZS}, Theorem 1]\label{theo:weight4}
Let $P(m)$ be an extended Preparata code and $\C_{P(m)}$ the extended
$1$-perfect code containing it. Then, $\C_{P(m)}=P(m)\cup Z(P(m))$, where $Z(P(m))=\{z\in\Z_2^n\,|\,d(c,z)\geq 4, \forall c\in P(m)\}$.
\end{theo}

\begin{coro}Let $P(m)$ be an extended Preparata code and $\C_{P(m)}$ the extended $1$-perfect code containing
it. Then, every word in $\C_{P(m)}$ is either in $P(m)$, or at distance $4$ from a word in $P(m)$.
\end{coro}

\begin{coro}\label{coro:blockInPrep}
Let $P(m)$ be an extended Preparata code, $\C_{P(m)}$ the extended
$1$-perfect code containing it and $SQS(n)$ the  Steiner quadruple
system corresponding to the minimum weight codewords of
$\C_{P(m)}$. Let $b$ be a $4$-tuple in $\{1,\dots,n\}$. Then,
either $b$ is a block in $SQS(n)$ and $b$ is not included in the
support of any codeword of weight $6$ in $P(m)$ or $b$ is included
in the support of exactly one codeword of weight $6$ in $P(m)$.
\end{coro}

\begin{proof}
Let $b$ be a $4$-tuple in $\{1,\dots,n\}$. If $b$ is a block in $SQS(n)$, then $c(b)$ is a codeword in $\C_{P(m)}$ and, by Lemma \ref{lemma:blockInPrep}, $b$ is not included in the support of any codeword of weight $6$ in $P(m)$. Assume $b$ is not a block in $SQS(n)$. Then, $c(b)$ is not a codeword in $\C_{P(m)}$ and, by Theorem \ref{theo:weight4}, there exists a codeword $c$ in $P(m)$ such that $d(c,c(b))\leq 3$. As the weight of such $c$ is at least $6$, then, necessarily $c$ is a codeword of weight $6$, $d(c,c(b))=2$, and hence $b\subset supp(c)$. Finally, if there exist $c'\in P(m)$ of weight $6$ such that $b\in supp(c')$, then $d(c,c')\leq 4$ that is not possible i $c\not=c'$.
\end{proof}

The question of whether the minimum
distance graph of a code uniquely determines it up to equivalence
first  arose \cite{PV} in the context of enumerating  and
isomorphism testing for  $1$-perfect codes \cite{Ph}. Later it
was solved in \cite{AZ} and \cite{MOOS} and used in subsequent
studies (e.g.\cite{Os}). In this article, we will establish that
the minimum distance graph of extended Preparata codes $P(m)$
uniquely determines the code up to equivalence. Alternatively, if
the distance graphs of two extended Preparata codes are
isomorphic, the codes are necessarily isometric and equivalent.
This result could prove useful in subsequent enumeration studies
as well.

  In Section \ref{section:weight6},
we will identify  the maximum size cliques in $DG$ and show that they
correspond to triples of coordinates. We will also identify sets
of maximum size cliques that correspond to pairs of coordinates. The
maximum  cliques will be labeled in Section
\ref{section:reconstruction} allowing for the labeling of all
vertices corresponding to codewords of minimum weight and,
eventually, all the vertices in the graph. Finally, conclusions
are given in Section \ref{section:conclusions}.
%We identify vertices of $DG$ with subsets or tuples of the set $\{1,\dots,n\}$ such that for any vertex $v$ the corresponding codeword in $P(m)$, denoted by $c(v)$, is the codeword whose support is the set $v$.

\section{Identification of all the weight $6$ codewords in $P(m)$}\label{section:weight6}

Let $DG$ be the minimum distance graph of an extended Preparata code $P(m)$. Identify one vertex $u_0\in DG$ as the all-zero codeword, $c(u_0)=\zero$. Consider $N(u_0)$ the set of neighbors of $u_0$. It will correspond to all
codewords of weight $6$ in the code. Two vertices $u,v\in N(u_0)$ are adjacent if and only if $|u\cap v|=3$.

\begin{lemm}\label{lemma:MaxInC}
Let $C$ be a clique in $\n$ and let $v=\{v_1,v_2,v_3,v_4,v_5,v_6\}\in C$.
\begin{enumerate}
\item[(i)] If there is a vertex $u$ in $C$ such that $|u\cap\trip|\leq 1$, then there are at most $2$ vertices in $C$ containing $\{v_1,v_2,v_3\}$.
\item[(ii)] If there is a vertex $u$ in $C$ such that $|u\cap\trip|=2$, then there are at most $4$ vertices in $C$ containing $\trip$.
\end{enumerate}
\end{lemm}

\begin{proof}
If $|u\cap\trip|\leq 1$, then any vertex, apart from $v$,
containing $\trip$ has to have intersection at least two with the
triple $u\setminus\{u\cap v\}$, and therefore there is only one
such vertex.

In the case $|u\cap\trip|=2$, then any vertex different to $v$ containing $\trip$ also has intersection $1$ with
the triple $u\setminus\{u\cap v\}$ and, hence, there are at most three of such vertices.
\end{proof}

\begin{prop}\label{prop:OtherCliques}
Let $C$ be a clique in $\n$ such that there is no triple $t$ intersecting all the vertices of $C$. Then, $|C|\leq 13$.
\end{prop}

\begin{proof}
Let $v=\vx\in C$. Consider $T$ the set of triples $\{u\cap v
\,|\,u\in C\}$. First, assume there are triples in $T$ that do not
have any element in common; for example, $\trip$ and
$\{v_4,v_5,v_6\}$. By Lemma \ref{lemma:MaxInC}, there is, apart from $v$, at most
one codeword containing each triple and, as the
distance between them is $6$, then both vertices have to be
$\{v_1,v_2,v_3,v_7,v_8,v_9\}$ and $\{v_4,v_5,v_6,v_7,v_8,v_9\}$.
Any other triple in $T$ intersects either $\trip$ or
$\{v_4,v_5,v_6\}$ in two elements. If it intersects $\trip$ in two
elements, then it intersects $\{v_7,v_8,v_9\}$ in one element and,
hence it has to intersect $\{v_4,v_5,v_6\}$ in two elements that
is a contradiction. In the same way, there is a contradiction if
it intersects in two elements with $\{v_4,v_5,v_6\}$ and, hence,
there is no other vertex in the clique and there are at most $3$
vertices in the clique.

Assume now that every two triples in $T$ have intersection. Let $N$ be the number of triple in $T$ that
appear in more than two vertices in $C$. We will study four different cases depending on the number $N$.
\begin{enumerate}
\item[(i)] $N=0$.\\
If $t \in T$ then $\bar{t}=v\setminus t$ can not be in $T$. If
$\bar{T}=\{ \bar{t}\, |\, t \in T \}$, then $|T \cup \bar{T}| \leq
20$, and $|T|=|\bar{T}|$, thus $|T|\leq 10$. The clique $C$ then
has at most $1+10=11$ vertices.

\item[(ii)] $N=1$.\\
If the triple $t$ appears in more than two vertices, then, by
Lemma \ref{lemma:MaxInC},
 every other triple in $T$ has intersection two with $t$. So, the number of triples in $T$ is at most ${3\choose 2}{3\choose 1}=9$ and the number of vertices in $C$ is at most $1+3+9=13$.\\

\item[(iii)] $N=2$.\\
Assume $t_1$ and $t_2$ are these two triples. By Lemma \ref{lemma:MaxInC}, the intersection of them is two, and
all the other triples in $T$ have intersection two with the both triples. Therefore, the number of such triples
is at most $4$, $2$ containing $t_1\cap t_2$ and $2$ containing only one element of $t_1\cap t_2$. Hence, the
number of vertices in $C$ is $1+2\cdot 3+4=11$.

\item[(vi)] $N\geq3$.\\
Consider $t_1,t_2$ and $t_3$, three triples included in more than two vertices. Any pair of them have two
elements in common (Lemma \ref{lemma:MaxInC}), and they also intersect in two elements with any other triple in
$T$. If $|t_1\cap t_2\cap t_3|=2$, then there is only one triple having intersection with them and also
contains $t_1\cap t_2\cap t_3$. If $|t_1\cap t_2\cap t_3|=1$, then the only triple intersecting the three of
them is $(t_1\cup t_2\cup t_3)\setminus (t_1\cap t_2\cap t_3)$. In both cases, the number of triples in $T$ is
$4$ and the new triple intersect with all the other triples of $T$ in two elements. Therefore, if there are more
than three triples in more than two vertices in $C$, then there are exactly $4$ of such triples and the number
of vertices in the clique is $1+4\cdot 3=13$.
\end{enumerate}

If we consider all the cases, then the maximum number of vertices in the clique is $max\{3,11,13,11,13\}=13$.

\end{proof}

\begin{prop}
If $DG$ is the minimum distance graph of an extended Preparata code of length $n=2^m$, $m\geq 6$, m even, then the
cliques of maximum size in $\n$ correspond to cliques $C(\trip)$, for $\trip\subseteq\{1,\dots,n\}$.
\end{prop}

\begin{proof}
Let $\trip$ be a triple in $\{1,\dots,n\}$. By Corollary \ref{coro:blockInPrep}, the $4$-set $\{v_1,v_2,v_3,X(\trip)\}$
is not included in the support of any codeword of $P(m)$ and, therefore, $X(\trip)$ is not included in any
vertex of $C(\trip)$. Moreover, since every $4$-tuple is included in exactly one vertex (Corollary \ref{coro:blockInPrep}), the clique $C(\trip)$ contains $\frac{n-4}{3}$ vertices. If $m\geq 6$, then $C(\trip)$ contains at least $20$ vertices and, by Proposition \ref{prop:OtherCliques}, it is a maximum size clique.
\end{proof}

Having established a one-to-one correspondence between triples and
maximum size cliques in $DG$, we now proceed to identify all triples
having a pair in common.

\begin{lemm} \label{lemma:neighbors}
Let $t_1$ and $t_2$ be triples in $\{1,\dots,n\}$ and consider $v\in C(t_1)$. If $|v\cap t_2|=1$, then $v$ has at most two neighbors in $C(t_2)$.
\end{lemm}

\begin{proof}
Since $|v\cap t_2|=1$, $v^{\prime}=\{v\setminus\{v\cap t_2\}\}$ is
a $5$-tuple in $\{1,\dots,n\}$. Any neighbor of $v$ in $C(t_2)$
contains two elements of $v^{\prime}$ and, therefore, there are at
most two of such neighbors.
\end{proof}

\begin{prop}
 Let $t_1$ and $t_2$ be triples in $\{1,\dots,n\}$, $n\geq 16$. If $C(t_1)$ and $C(t_2)$
 have
 no vertex in common and $|t_1\cap t_2|\leq 1 $, then there is a vertex in one of the cliques that
 has less than two neighbors in the other clique.
\end{prop}

\begin{proof}
If $|t_1\cap t_2|= 1$ and $n\geq 16$, then there exist a triple
$t$ such that $|t\cap t_2|=0$ and $t\cup t_1\in C(t_1)$. Then, by
Lemma \ref{lemma:neighbors}, $t\cup t_1$ has at most two neighbors
in $C(t_2)$.

Assume $t_1=\{v_1,v_2,v_3\}, t_2=\{v_4,v_5,v_6\}$, $|t_1\cap t_2|=
0$. As $C(t_1)$ and $C(t_2)$ have no vertex in common, $t_1\cup
t_2$ is not a vertex in $DG$. If there is one vertex in $C(t_1)$
with exactly one element of $t_2$, then such vertex has two
neighbors in $C(t_2)$ by Lemma \ref{lemma:neighbors}. The same
argument  applies to clique $C(t_2)$. Otherwise, there is a vertex
$u_1=t_1 \cup s_1$ in $C(t_1)$ with  $|s_1 \cap t_2|=2$,  and a
vertex $u_2=t_2\cup s_2$ in $C(t_2)$ with two elements of
$|s_2\cap t_1|=2$. In that case, $|u_1\cap u_2|= 4$ and
therefore $d(c(u_1),c(u_2))<6$ which is not possible.
%with all other vertices having no intersect with $t_2$
%($t_1$ respectively). For example, we can assume $\{v_1,v_2,v_3,v_4,v_5,x\}= u_1$, and
%$\{v_4,v_5,v_6,v_1,v_2,y\}= u_2$, but then $d(c(u_1),c(u_2))<6$ which is not possible.
\end{proof}

\begin{prop}\label{prop:4-partite}
Let $t_1$ and $t_2$ be triples in $\{1,\dots,n\}$,  $n\geq 16$, such that $C(t_1)$ and $C(t_2)$ have no vertex in
common. If every vertex in $C(t_1)$ has $3$ neighbors in $C(t_2)$ and vice versa, then $|t_1\cap t_2|=2$ and
$t_1\cup t_2$ is a block in $SQS(n)$.
\end{prop}

\begin{proof}
Assume $C(t_1)$ and $C(t_2)$ have no vertex in common. If $|t_1\cap t_2|\leq 1$ then there exists one vertex in
one of  the cliques with less that $3$ neighbors in the other clique. Hence, if every vertex in each clique has
$3$ neighbors in the other, then $|t_1\cap t_2|=2$. If $t_1\cup t_2$ is not a block in $SQS(v)$, then the vertex
containing $t_1\cup t_2$ belongs to the intersection of both cliques that is not possible.
\end{proof}

\begin{coro}\label{coro:4-partite}
There is a one-to-one correspondence between  blocks in $SQS(n)$ and  $4$-partite $3$-regular graphs in the minimum
distance graph of an extended Preparata code.
\end{coro}

\begin{proof}
Let $t_1,t_2$ be triples such that $t_1 \cup t_2 =b=\{v_1,v_2,v_3,v_4\}\in SQS(n)$. Since $b$ is not contained
in any vertex of $DG$ (Lemma \ref{lemma:blockInPrep}), $C(t_1)$ and $C(t_2)$ have no vertex in common.  Consider
the vertex $v=t_1\cup\{v_5,v_6,v_7\}\in C(t_1)$. Then, by Corollary \ref{coro:blockInPrep}, the $4$-sets $t_2\cup\{v_5\}$, $t_2\cup\{v_6\}$ and
$t_2\cup\{v_7\}$ must belong to three different vertices in $C(t_2)$ and, therefore, $v \in C(t_1)$ has three
neighbors in $C(t_2)$. The converse is given by Proposition \ref{prop:4-partite}.
\end{proof}

\begin{prop}\label{prop:Cab}
Let $t_1$ and $t_2$ be triples in $\{1,\dots,n\}$, $n\geq 16$,
such that $C(t_1)$ and $C(t_2)$ have one vertex in common. There
is a vertex in each clique with $2$ neighbors in the other clique
and every other vertex in each clique, apart from the
intersection, has $3$ neighbors in the other clique if and only if
$|t_1\cap t_2|=2$.
\end{prop}

\begin{proof}
Since $C(t_1)$ and $C(t_2)$ have one vertex in common, $X(t_1)$
and $X(t_2)$ appear in $C(t_2)$ and $C(t_1)$ respectively. Assume
$t_1$ and $t_2$ has $2$ elements in common. The vertex in $C(t_1)$
containing $t_1\cup X(t_2)$ has exactly $2$ neighbors in $C(t_2)$
and the vertex in $C(t_2)$ containing $t_2\cup X(t_1)$ has $2$
neighbors in $C(t_1)$. Any other vertex, apart from the vertex in
the intersection of the cliques, has exactly $3$ neighbors in the
other clique.

If $|t_1\cap t_2|=1$, then every vertex $v$ in $C(t_1)$ with $3$ neighbors in $C(t_2)$ has intersection with the pair $t_2\setminus(t_1\cap t_2)$, by Lemma \ref{lemma:neighbors}, but it is not possible if there are more than $2$ vertices with $3$ neighbors, that is the case if $n\geq 16$.

Finally, if $t_1$ and $t_2$ do not have intersection, then every vertex in each clique have, apart from the vertex in the intersection, at most one neighbor in the other clique.
\end{proof}

Define the set of all the cliques determined by triples with two elements in common:
\[
S(\{v_1,v_2\})=\{C(t)\,|\, |t|=3, \{v_1,v_2\}\subset t\}
\]

\begin{coro}\label{coro:Sxy}
Given $\n$, we can identify all the sets $S(t)$, where $t$ is a pair in $\{1,\dots,n\}$.
\end{coro}
\begin{proof}
Propositions \ref{prop:4-partite} and \ref{prop:Cab}.
\end{proof}

\section{Reconstruction of the extended Preparata code}\label{section:reconstruction}

From last section, we have fixed a vertex $u_0$ as the all-zero codeword and we have determined the set of
vertices corresponding to codewords of weight 6. Moreover, we can identify all cliques of maximum size with
triples. In order to reconstruct the extended Preparata code from its minimum distance graph, we will label each vertex in the graph, that is, we will associate a subset $v$ of $\{1,\dots,n\}$ such that the corresponding codeword in the code will be $c(v)$. Since cliques of maximum size correspond to cliques of type $C(t)$, where $t$ is a triple in $\{1,\dots,n\}$, then by labeling a clique $C$ we will mean associate a triple $t$ in $\{1,\dots,n\}$ such that $C=C(t)$. Similarly, to label a set $S(p)$ is to determine the pair $p$.

We consider one clique of maximum size and we label it as
$C(\{1,2,3\})$. We define $X(\{1,2,3\})=n$ and choose disjoint
triples $t$ of $\{4,\dots,n-1\}$. We then label all the vertices
in $C(\{1,2,3\})$ as $\{1,2,3\}\cup t$. In Subsection
\ref{subsection:CliquesInN} we will label first all the maximum
size cliques having intersection with $C(\{1,2,3\})$ and after
that any maximum size clique in $\n$ and, hence, all codewords in
$\n$. Finally, in Subsection \ref{subsection:CliquesInDG} we will
label all cliques of maximum size in the distance graph and
therefore all the codewords of the extended Preparata code.

\subsection{Identification of the maximum size cliques in $\n$}\label{subsection:CliquesInN}
Consider the clique $C(\{1,2,3\})$. By Corollary \ref{coro:Sxy}, we can identify three sets $S(p)$ containing
$C(\{1,2,3\})$, where $p$ is a pair, $p\subset\{1,2,3\}$. We label these sets as $S(\{1,2\})$, $S(\{2,3\})$ and
$S(\{1,3\})$.

Let $v=\{1,2,3,v_4,v_5,v_6\}$ be a vertex in $C(\{1,2,3\})$. There are ${6\choose 3}=20$ different triples
$t\subset v$ such that $v\in C(t)$ and, therefore, $20$ cliques $C(t)$ intersecting $C(\trip)$ in the vertex
$v$.

Consider the set $S(\{1,2\})$. There are three maximum size cliques, apart from $C(\{1,2,3\})$, that intersect
$C(\{1,2,3\})$ in $v$ and are included in $S(\{1,2\})$. We label them as $C(\{1,2,v_4\})$, $C(\{1,2,v_5\})$ and
$C(\{1,2,v_6\})$. The clique $C(\{1,2,v_4\})$ belong to the $3$ sets $S(\{1,2\})$, already labeled,
$S(\{1,v_4\})$ and $S(\{2,v_4\})$. The set $S(\{1,v_4\})$ is the one that intersects with the set $S(\{1,3\})$
and the clique in the intersection is $C(\{1,3,v_4\})$. Similarly, we also label $S(\{2,v_4\})$ that intersects with $S(\{2,3\})$ and the intersection is $C(\{2,3,v_4\})$.

Doing the same process with the cliques  $C(\{1,2,v_5\})$ and $C(\{1,2,v_6\})$ we label all the sets $S(\{i,k\})$ and the $9$ cliques $C(\{i,j,k\})$, where $i,j\in\{1,2,3\}$ and $k\in\{v_4,v_5,v_6\}$.

In order to label the cliques $C(\{i,k,l\})$, where $i\in\{1,2,3\}$ and $k,l\in\{v_4,v_5,v_6\}$, we consider the intersection on the sets $S(\{i,k\})$ and $S(\{i,l\})$ that are already labeled. So, we can also label the sets $S(\{k,l\})$, where $k,l\in\{v_4,v_5,v_6\}$ and finally the clique $C(\{v_4,v_5,v_6\})$.

That way the $20$ cliques containing $v$ are labeled, and we can repeat the same process for any vertex in
$C(\{1,2,3\})$.

\begin{prop}\label{prop:AllCliques}
Let $x$ be a vertex in the minimum distance graph $DG$ of an extended Preparata code and $N(x)$ the set of neighbors of
$x$. Let $C$ be a clique of maximum size in $N(x)$. If all the vertices in $C$ and all the maximum size cliques
intersecting $C$ are labeled, then all the vertices in $N(x)$ are determined.
\end{prop}

\begin{proof}
Assume all the vertices of $C$ and all the cliques of maximum size intersecting $C$ are labeled. That way, all
the cliques $C(t)$ are labeled, where $t$ is a triple contained in some vertex of $C$. Moreover, we can identify
and label all the sets $S(p)$, where $p$ is a pair included in some vertex of $C$. Consider the triple $\trip$
such that $C=C(\trip)$. Let $t'=\{v_i,v_4,v_5\}$ be a triple where $v_i\in\trip$ and $v_4$ and $v_5$ belong to
different vertices in $C$. Then, $S(\{v_i,v_4\})$ and $S(\{v_i,v_5\})$ are labeled and
$C(\{v_i,v_4,v_5\})=S(\{v_i,v_4\})\cap S(\{v_i,v_5\})$. After labeling all the cliques of this type, we can
identify the sets $S(p)$ where $p$ is any pair in $\{1,\dots,n\}\setminus X(\trip)$. Hence, for any triple
$\{i,j,k\}\in\{1,\dots,n\}\setminus X(\trip)$, $C(\{i,j,k\})=S(\{i,j\})\cap S(\{j,k\})$. Finally, for any pair
$p$ in $\{1,\dots,n\}\setminus X(\trip)$, all the maximum size cliques in $S(p)$ are labeled except one, that is
the clique $C(p\cup X(\trip)$ and, therefore all the maximum size cliques are labeled and, hence, all vertices
in $N(x)$.
\end{proof}

\subsection{Identification of all the vertices in the distance graph}\label{subsection:CliquesInDG}

In last subsection we have labeled all the vertices in $DG$ corresponding to weight $6$ codewords in the
extended Preparata code. For $u\in DG$, denote $N(u)$ be the set of neighbors of $u$, $N(u)^+=N(u)\cup u$ and $c(N(u))$ the subcode $\{c(v)\,|\,v\in N(u)\}$ of $P(m)$. Then, we define $N(u)+u$ as $\{supp(v)\,|\, v \in c(N(u))+u \}$.

Let $u=\vx$ be one of vertex corresponding to weight $6$ codewords in the
extended Preparata code and $\bar{u}$ the set $\{1,\dots,n\}\setminus u$. Note that all the codewords in $c(N(u))+u$ have
weight $6$ and their supports will be subsets of size $6$ of $\{1,\dots, n\}$. We will label $N(u)+u$ and,
therefore, it will be labeled $N(u)$.

The vertex $u$ is included in $C(\trip)\subset N(u_0)^+\cap N(u)^+$. That clique is labeled as $C(\{v_4,v_5,v_6\})$
in $N(u)+u$ and all the vertices $\{v_1,v_2,v_3,v_7,v_8,v_9\}\in C(\trip)\subset N(u_0)^+\cap N(u)^+$ are also
labeled as $\{v_4,v_5,v_6,v_7,v_8,v_9\}$ in $C(\{v_4,v_5,v_6\})\subset N(u)+u$. Moreover, any maximum size
clique $C(t)\subset N(u_0)^+\cap N(u)^+$, where $t$ is a triple, is labeled in $N(u)+u$ and also all its vertices.

If we identify $S(p)$, for any pair $p\subset\{1,\dots,n\}$, then we will have labeled all vertices in $N(u)+u$.
Since all the maximum size cliques $C(t)$ where $t\in u$ are labeled, we can identify the sets $S(p)$, where $p$ is a pair in $u$. In order to identify $S(\{v_i,v_s\})$ where
$v_i\in u$, $v_s\in\bar{u}$, consider $v_j\in u$, $v_j\not = v_i$. For all $v_k\in u \setminus\{v_i,v_j\}$, the vertex containing
$\{v_i,v_j,v_k,v_s\}$ is included in $C(\{v_i,v_j,v_k\})$ and, hence it is already labeled. That way,
$\{v_i,v_j,v_k,v_s\}$, for $v_k\in u\setminus\{v_i,v_j\}$,  are vertices included in a maximum size clique that
correspond to $C(\{v_i,v_j,v_s\})$. Hence, we can label all the maximum size cliques $C(\{v_i,v_j,v_s\})$, for $v_i,v_j\in u$, $v_s \in \bar{u}$ and therefore, all the sets $S(\{v_i,v_s\})$ where $v_i\in u$,
$v_s\in\bar{u}$. Cliques $C(\{v_i,v_j,v_s\})$, where
$v_i\in u$, $v_j,v_s\in\bar{u}$ can be identified by $S(\{v_i,v_s\})\cap S(\{v_i,v_j\})$. Finally, with last cliques labeled, we identify $S(\{v_i,v_j\})$,
$v_i,v_j\in\bar{u}$, and $C(\{v_i,v_j,v_s\})=S(\{v_i,v_s\})\cap (\{v_j,v_s\})$ where $v_i,v_j,v_s\in\bar{u}$.

In $N(u)+u$ we have labeled the clique $C(\trip)$, all the vertices in the clique and all the maximum size cliques intersecting $C(\trip)$. Hence, by Proposition \ref{prop:AllCliques}, we can identify all the vertices in $N(u)+u$ and, therefore, in $N(u)$. We can repeat this process with any vertex in $\n$ and any new vertex labeled in the graph and we obtain all the vertices labeled in the minimum distance graph $DG$.

\section{Conclusions}\label{section:conclusions}

As a consequence of the previous arguments, we have established the following results.

\begin{theo}
The minimum distance graph of extended Preparata codes $P(m)$
uniquely determines the code up to equivalence. Alternatively, the
distance graphs of two extended Preparata codes are isomorphic if
and only if the codes are  equivalent.
\end{theo}

\begin{coro}
Given the minimum distance graph for $P(m)$, one can reconstruct
 the corresponding extended perfect code
$C_{P(m)}$ and its  distance $4$ graph.
\end{coro}

One could also ask what relation, if any, there is between the
minimum distance graph of $P(m)$ and that of $C_{P(m)}$.

\end{document}